\documentclass[twoside,a4paper,11pt]{sca}
\usepackage{graphicx}
\usepackage{hyperref}
\usepackage{movie15}
\usepackage{natbib}  
\begin{document}
\pagenumbering{arabic}
\pagestyle{myheadings}
\thispagestyle{empty}
{\flushright\includegraphics[width=\textwidth,bb=90 650 520 700]{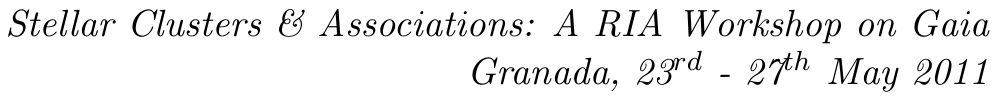}}
\vspace*{0.2cm}
\begin{flushleft}
{\bf {\LARGE
%
An Aladin-based search for proper-motion companions to young stars in the Local Association, Tucana-Horologium and $\beta$\,Pictoris
%
}\\
\vspace*{1cm}
%
F. J. Alonso-Floriano$^{1}$,
J. A. Caballero$^{2}$, 
and 
D. Montes$^{1}$
%
}\\
\vspace*{0.5cm}
%
$^{1}$
Departamento de Astrof\'isica y Ciencias de la Atm\'osfera, Facultad de F\'isica, Universidad Complutense de Madrid, E-28040 Madrid, Spain\\
$^{2}$
Centro de Astrobiolog\'ia (CSIC-INTA), PO Box 78, E-28691 Villanueva de la Ca\~nada, Madrid, Spain
%
\end{flushleft}
%
\markboth{
An Aladin-based search for young proper-motion companions
}{ 
%
Alonso-Floriano, F. J.  et al.
%
}
\thispagestyle{empty}
\vspace*{0.4cm}
\begin{minipage}[l]{0.09\textwidth}
\ 
\end{minipage}
\begin{minipage}[r]{0.9\textwidth}
\vspace{1cm}
\section*{Abstract}{\small
%
We have used the Aladin sky atlas of the Virtual Observatory to look for new common proper-motion pairs in three young stellar kinematic groups: 
Local Association ($\tau \sim$ 10--120\,Ma), Tucana-Horologium ($\tau \sim$ 30\,Ma) and $\beta$ Pictoris ($\tau \sim$ 12\,Ma). 
We have found 9 new and 14 known common proper-motion companions to the 210 investigated stars. 
With the CAFOS instrument at the 2.2\,m Calar Alto telescope, we have investigated in detail one of the new pairs, the HD 143809 AB system, which is formed by a bright G0V primary star and a previously unknown young M1.0--1.5Ve star.  
%
\normalsize}
\end{minipage}
%
%
%

\section{Introduction \label{section.intro}}

Young nearby late-type stars are excellent targets for high-contrast imaging surveys for brown dwarf and planetary companions.
Many of these young low-mass stars in the solar neighbourhood belong to stellar kinematics groups with ages younger than the Pleiades, such as the Local Association and its kinematic subgroups \citep{Montes01b, Song03, Zuckerman04}.
One way of identifying such stars is searching for faint proper-motion companions at wide separations to already-known members in young stellar kinematic groups.

We followed the procedure described by \cite{Caballero10a} 
and used a powerful Virtual Observatory tool, the Aladin sky atlas \citep{Bonnarel00}, to look for proper-motion companions to stars in the Local Association (LA, $\tau \sim$ 10--120\,Ma), Tucana-Horologium (Tuc-Hor, $\tau \sim$ 30\,Ma) and $\beta$\,Pictoris ($\beta$\,Pic, $\tau \sim$ 12\,Ma) stellar kinematic groups.  

\begin{table}[]
	\caption{\label{tab.1}The nine unknown proper-motion pairs (SKG indicates the stellar kinematic group).}
	\center
	\small
	\begin{tabular}{ l c c c c c}
	\hline\hline
Name				&	Sp.			&	$\rho$			&	$\theta$				&	$s$				&	SKG			\\
					&	type			&	[arcsec]			&	[deg]					&	[kAU]			&				 \\ [0.5ex]   
	\hline 
HD 82939 			&	G5V			&	162.28 $\pm$ 0.17	&	121.49 $\pm$ 0.07		&	6.3 $\pm$ 0.2		&	LA			\\
GJ 9303 				&	K7V			&					&						&					&				\\
\noalign{\smallskip}
EX Cet				&	G5V			&	612.10 $\pm$ 0.11	&	258.66 $\pm$ 0.02		&	14.7 $\pm$ 0.3		&	LA			\\
G 271--110 			&	M3.5V		&					&						&					&				\\
\noalign{\smallskip}
HD 143809 A 			&	G0V			&	86.40 $\pm$ 0.11	&	252.57 $\pm$ 0.09		&	7.1 $\pm$ 0.9		&	LA			\\
HD 143809 B 			&	M1.0--1.5V	&					&						&					&				\\
\noalign{\smallskip}
HD 13183 			&	G7V			&	705.99 $\pm$ 0.10	&	103.761 $\pm$ 0.009	&	36.0 $\pm$ 1.2		&	Tuc-Hor		\\
CD--53 413			&	G5V			&					&						&					&				\\
\noalign{\smallskip}
CD--53 544 			&	K6Ve		&	22.06 $\pm$ 0.08	&	11.11 $\pm$ 0.17		&	0.93 $\pm$ 0.09	&	Tuc-Hor		\\
AF Hor 				&	M2Ve		&					&						&					&				\\
\noalign{\smallskip}
HD 207964 AB			&	F1III+...		&	1412.75 $\pm$ 0.11	&	245.020 $\pm$ 0.005	&	64.0 $\pm$ 1.9		&	Tuc-Hor		\\
HD 207575 			&	F6V			&					&						&					&				\\
\noalign{\smallskip}	
HD 173167 			&	F5V	 		&	550.31 $\pm$ 0.10	&	 290.244 $\pm$ 0.011	&	29 $\pm$ 3		&	$\beta$\,Pic	\\	
TYC 9073--0762--1 		&	M1Ve		&					&						&					&				\\
\noalign{\smallskip} 
$\eta$ Tel	 AB			&	A0Vn+M7.5V	&	416.26 $\pm$ 0.13	&	170.691 $\pm$ 0.012	&	20.1 $\pm$ 0.2		&	$\beta$\,Pic	\\
HD 181327 			&	F6V			&					&						&					&				\\
\noalign{\smallskip}
HD 199143 AB		 	&	F7V+...		&	325.04 $\pm$ 0.08	&	138.35 $\pm$ 0.02		&	14.8 $\pm$ 0.5		&	$\beta$\,Pic	\\
AZ Cap 				&	K6Ve		&					&						&					&				\\ [0.5ex]
	\hline
	\end{tabular} 
	\label{table.unknown}
\end{table}

\section{Analysis \label{section.analysis}}

We searched for either primary (i.e., brighter) and secondary (i.e., fainter) companions to 210 nearby young stars in the three different moving groups compiled by \cite{Montes01b} and \cite{Torres08}: 116 in LA, 44 in Tuc-Hor, and 50 in $\beta$\,Pic. We used the interactive software Aladin v5 to load 2MASS \citep{Skrutskie06} and USNO-B1 \citep{Monet03} astro-photometric catalogues and cross-matched them in a circular area of radius 30\,arcmin centred on each target. 
Next, we constructed a proper-motion diagram with the Aladin application VOPlot and searched for sources with USNO-B1 proper motions different from those of the target stars by less than 10\,mas\,a$^{-1}$. 

For each proper-motion candidate, we looked for previous claiming in the literature of membership in multiple systems (e.g., the Washington Double Star catalogue; \citealt{Mason01}) and for better proper-motion determinations (e.g., Tycho-2; \citealt{Hog00}).
We derived photometric distances for both primaries and secondaries, based on available spectroscopy or photometry, of the candidate pairs without parallactic distance measurements and discarded those with no coincident values.

\section{Results}

\begin{figure}[!h]
\center
\includegraphics[width=0.99\textwidth]{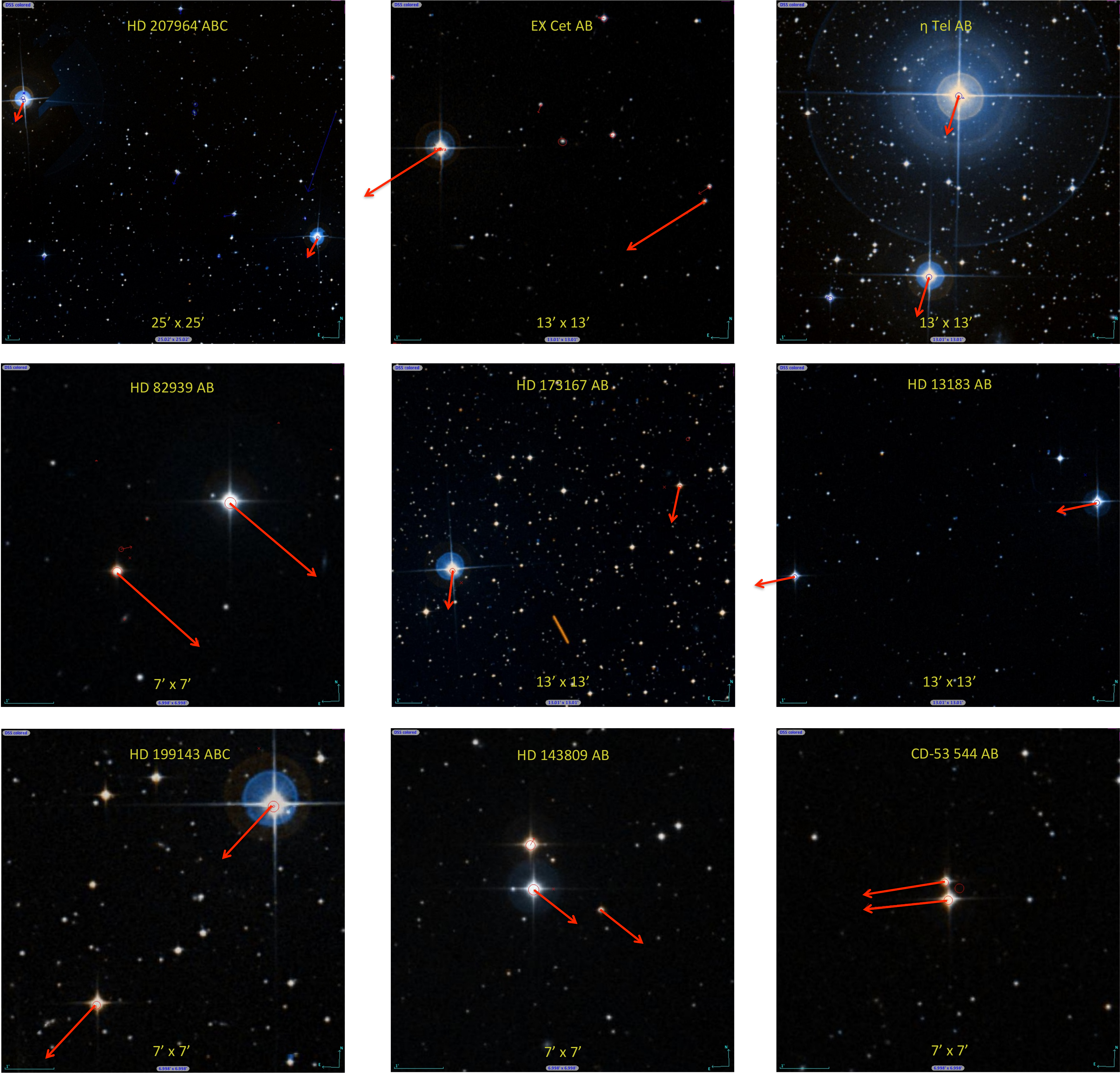} 
\caption{\label{fig.9pairs} False-colour images combining DSS POSSII $B_J$, $R_F$ and $I_N$ photographic plates of the nine new pairs.
Labelled are the multiple system names and field-of-view sizes.
North is up and east is to the left.
The red arrows show the proper motion.}
\end{figure}

Of the 210 investigated stars, we identified 23 multiple system candidates, of which 14 were known common proper-motion companions and 9 were unknown multiple systems (see Table~\ref{tab.1} and Figure~\ref{fig.9pairs}). One of the nine of them was a suspected multiple system ($\eta$~Tel~AB and HD\,181327; \citealt{Schneider06}).
At the measured or derived distances, the angular separations of 0.37 to 24\,arcmin translate into projected physical separations between 0.0045 and 0.31\,pc.
Interestingly, some of the proper-motion companions had been already tabulated as members in the same stellar kinematic group as their target stars (e.g., HD 207964 AB and HD 207575 in Tucana-Horologium).

One of the new multiple systems was formed by a young solar analogue and an anonymous high proper-motion red dwarf never described in the literature before, and  was subject of a dedicated astrometric, photometric and spectroscopic follow-up study.
First, we confirmed the common proper motion of the ``HD\,143809\,AB'' system using 11 astrometric epochs separated by over 56 years as in \cite{Caballero10b}. 
Next, we collected $B$, $V$, $R$ and $I$ images and low-resolution spectra (grating G100) with the CAFOS instrument at the 2.2\,m Calar Alto telescope.
While the primary is a known G0V star with a high lithium abundance ($EW$(Li~{\sc i}) = 103\,m{\AA};  \citealt{Lopez-Santiago10}), an estimated age of $\tau \sim$ 80--120\,Ma and kinematics consistent with membership in the Local Association \citep{Montes01a}, the new companion at $s$ = 7.1 $\pm$ 0.9\,kAU is an M1.0--1.5Ve star with chromospheric H$\alpha$, H$\beta$ and H$\gamma$ emission.
The heliocentric distance derived from its spectral type and photometry matches the one of the primary measured by {\em Hipparcos} at $d$ = 78 $\pm$ 8\,pc. 
Using this distance, its $J$-band apparent magnitud ($J$ = 10.35 $\pm$ 0.03\,mag) and the NextGen models \citep{Baraffe98} for an age of 100\,Ma, HD\,143809\,B has a most probable mass of 0.57--0.60\,$M_\odot$.

%



%
\small  
%
\section*{Acknowledgments}   
%
We thank J. Genebriera for providing us unpublished images of the HD\,143809\,AB system.
This research has made use of Aladin.
This work is supported by the Universidad Complutense de Madrid, the Spanish Ministerio de Ciencia e Innovaci\'on (MICINN) under grant AYA2008-0695, and the Comunidad de Madrid under PRICIT project S2009/ESP-1496 (AstroMadrid).
%
%
%
%
%

\bibliographystyle{aa}
\bibliography{Poster_Alonso-Floriano_FJ}

\begin{thebibliography}{15}
\expandafter\ifx\csname natexlab\endcsname\relax\def\natexlab#1{#1}\fi

\bibitem[{{Baraffe} {et~al.}(1998){Baraffe}, {Chabrier}, {Allard}, \&
  {Hauschildt}}]{Baraffe98}
{Baraffe}, I., {Chabrier}, G., {Allard}, F., \& {Hauschildt}, P.~H. 1998, A\&A,
  337, 403

\bibitem[{{Bonnarel} {et~al.}(2000){Bonnarel}, {Fernique}, {Bienaym{\'e}},
  {Egret}, {Genova}, {Louys}, {Ochsenbein}, {Wenger}, \&
  {Bartlett}}]{Bonnarel00}
{Bonnarel}, F., {Fernique}, P., {Bienaym{\'e}}, O., {et~al.} 2000, A\&A, 143,
  33

\bibitem[{{Caballero} {et~al.}(2010{\natexlab{a}}){Caballero}, {Miret},
  {Genebriera}, {Tobal}, {Cairol}, \& {Montes}}]{Caballero10a}
{Caballero}, J.~A., {Miret}, F.~X., {Genebriera}, J., {et~al.}
  2010{\natexlab{a}}, in Highlights of Spanish Astrophysics V, ed.
  {J.~M.~Diego, L.~J.~Goicoechea, J.~I.~Gonz{\'a}lez-Serrano, \& J.~Gorgas},
  379

\bibitem[{{Caballero} {et~al.}(2010{\natexlab{b}}){Caballero}, {Montes},
  {Klutsch}, {Genebriera}, {Miret}, {Tobal}, {Cairol}, \&
  {Pedraz}}]{Caballero10b}
{Caballero}, J.~A., {Montes}, D., {Klutsch}, A., {et~al.} 2010{\natexlab{b}},
  A\&A, 520, A91

\bibitem[{{H{\o}g} {et~al.}(2000){H{\o}g}, {Fabricius}, {Makarov}, {Urban},
  {Corbin}, {Wycoff}, {Bastian}, {Schwekendiek}, \& {Wicenec}}]{Hog00}
{H{\o}g}, E., {Fabricius}, C., {Makarov}, V.~V., {et~al.} 2000, A\&A, 355, L27

\bibitem[{{L{\'o}pez-Santiago} {et~al.}(2010){L{\'o}pez-Santiago}, {Montes},
  {G{\'a}lvez-Ortiz}, {Crespo-Chac{\'o}n}, {Mart{\'{\i}}nez-Arn{\'a}iz},
  {Fern{\'a}ndez-Figueroa}, {de Castro}, \& {Cornide}}]{Lopez-Santiago10}
{L{\'o}pez-Santiago}, J., {Montes}, D., {G{\'a}lvez-Ortiz}, M.~C., {et~al.}
  2010, A\&A, 514, A97

\bibitem[{{Mason} {et~al.}(2001){Mason}, {Wycoff}, {Hartkopf}, {Douglass}, \&
  {Worley}}]{Mason01}
{Mason}, B.~D., {Wycoff}, G.~L., {Hartkopf}, W.~I., {Douglass}, G.~G., \&
  {Worley}, C.~E. 2001, AJ, 122, 3466

\bibitem[{{Monet} {et~al.}(2003){Monet}, {Levine}, {Canzian}, {Ables}, {Bird},
  {Dahn}, {Guetter}, {Harris}, {Henden}, {Leggett}, {Levison}, {Luginbuhl},
  {Martini}, {Monet}, {Munn}, {Pier}, {Rhodes}, {Riepe}, {Sell}, {Stone},
  {Vrba}, {Walker}, {Westerhout}, {Brucato}, {Reid}, {Schoening}, {Hartley},
  {Read}, \& {Tritton}}]{Monet03}
{Monet}, D.~G., {Levine}, S.~E., {Canzian}, B., {et~al.} 2003, AJ, 125, 984

\bibitem[{{Montes} {et~al.}(2001{\natexlab{a}}){Montes}, {L{\'o}pez-Santiago},
  {Fern{\'a}ndez-Figueroa}, \& {G{\'a}lvez}}]{Montes01a}
{Montes}, D., {L{\'o}pez-Santiago}, J., {Fern{\'a}ndez-Figueroa}, M.~J., \&
  {G{\'a}lvez}, M.~C. 2001{\natexlab{a}}, A\&A, 379, 976

\bibitem[{{Montes} {et~al.}(2001{\natexlab{b}}){Montes}, {Lopez-Santiago},
  {Galvez}, {Fernandez-Figueroa}, {de Castro}, \& {Cornide}}]{Montes01b}
{Montes}, D., {Lopez-Santiago}, J., {Galvez}, M.~C., {et~al.}
  2001{\natexlab{b}}, MNRAS, 732, 45

\bibitem[{{Schneider} {et~al.}(2006){Schneider}, {Silverstone}, {Hines},
  {Augereau}, {Pinte}, {M{\'e}nard}, {Krist}, {Clampin}, {Grady}, {Golimowski},
  {Ardila}, {Henning}, {Wolf}, \& {Rodmann}}]{Schneider06}
{Schneider}, G., {Silverstone}, M.~D., {Hines}, D.~C., {et~al.} 2006, ApJ, 650,
  414

\bibitem[{{Skrutskie} {et~al.}(2006){Skrutskie}, {Cutri}, {Stiening},
  {Weinberg}, {Schneider}, {Carpenter}, {Beichman}, {Capps}, {Chester},
  {Elias}, {Huchra}, {Liebert}, {Lonsdale}, {Monet}, {Price}, {Seitzer},
  {Jarrett}, {Kirkpatrick}, {Gizis}, {Howard}, {Evans}, {Fowler}, {Fullmer},
  {Hurt}, {Light}, {Kopan}, {Marsh}, {McCallon}, {Tam}, {Van Dyk}, \&
  {Wheelock}}]{Skrutskie06}
{Skrutskie}, M.~F., {Cutri}, R.~M., {Stiening}, R., {et~al.} 2006, AJ, 131,
  1163

\bibitem[{{Song} {et~al.}(2003){Song}, {Zuckerman}, \& {Bessell}}]{Song03}
{Song}, I., {Zuckerman}, B., \& {Bessell}, M.~S. 2003, ApJ, 599, 342

\bibitem[{{Torres} {et~al.}(2008){Torres}, {Quast}, {Melo}, \&
  {Sterzik}}]{Torres08}
{Torres}, C.~A.~O., {Quast}, G.~R., {Melo}, C.~H.~F., \& {Sterzik}, M.~F. 2008,
  {Handbook of Star Forming Regions, Volume II: The Southern Sky, ASP Monograph
  Publications, Vol. 5}, ed. {Reipurth, B.}, 757

\bibitem[{{Zuckerman} \& {Song}(2004)}]{Zuckerman04}
{Zuckerman}, B. \& {Song}, I. 2004, ARA\&A, 42, 685

\end{thebibliography}

\end{document}